\documentclass[prd,superscriptaddress,noshowpacs,showkeys,twocolumn]{revtex4-1}
\pdfoutput=1

\usepackage{lineno}
\usepackage{amsfonts}
\usepackage{amsmath}
\usepackage{amssymb}
\usepackage{graphicx}
\usepackage{units}
\usepackage{siunitx}
\usepackage{multirow}
\usepackage{color}
\usepackage{bm}
\usepackage[colorlinks=true,linkcolor=blue]{hyperref}
\usepackage{epstopdf}
\usepackage{bigstrut}
\usepackage{xspace}
\usepackage{gensymb}
\usepackage{diagbox}

\RequirePackage{lineno}

\newcommand{\etap}{\eta^{\prime}}
\newcommand{\etasl}{\eta e^+\nu_{e}}
\newcommand{\etapsl}{\eta' e^+\nu_{e}}
\newcommand{\etappsl}{\etapp e^+\nu_{e}}
\newcommand{\dpetasl}{D^+\to\etasl}
\newcommand{\dpetapsl}{D^+\to\etapsl}
\newcommand{\dppsl}{D^+\to\etappsl}

\newcommand{\vcd}{V_{cd}}
\newcommand{\ee}{e^+e^-}

\newcommand{\ddbar}{D\bar{D}}

\newcommand{\gev}{\,\unit{GeV}}
\newcommand{\gevc}{\,\unit{GeV}/\emph{c}}
\newcommand{\gevcc}{\,\unit{GeV}/\emph{c}^2}
\newcommand{\gevccs}{\,\unit{GeV^2}/\emph{c}^4}
\newcommand{\br}[1]{\mathcal{B}_{#1}}
\newcommand{\ks}{K^0_S}

\newcommand{\mbc}{M_\mathrm{BC}}
\newcommand{\etapp}{\eta^{(\prime)}}
\newcommand{\delE}{\Delta\emph{E}}

\newcommand\Tstrut{\rule{0pt}{2.4ex}}       % top strut
\newcommand\Bstrut{\rule[-1.3ex]{0pt}{0pt}} % bottom strut

\begin{document}

%\pagewiselinenumbers
%\modulolinenumbers[2]
%\linenumbers
%\begin{frontmatter}

\title{\boldmath Study of the decays $D^+\rightarrow\etapp e^+\nu_{e}$}

\author{
\begin{small}
\begin{center}
M.~Ablikim$^{1}$, M.~N.~Achasov$^{9,d}$, S. ~Ahmed$^{14}$, M.~Albrecht$^{4}$, A.~Amoroso$^{53A,53C}$, F.~F.~An$^{1}$, Q.~An$^{50,40}$, Y.~Bai$^{39}$, O.~Bakina$^{24}$, R.~Baldini Ferroli$^{20A}$
% ...
\\
\vspace{0.2cm}
(BESIII Collaboration)\\
\vspace{0.2cm} {\it
$^{1}$ Institute of High Energy Physics, Beijing 100049, People's Republic of China\\
% ...
$^{4}$ Bochum Ruhr-University, D-44780 Bochum, Germany\\
% ...
$^{9}$ G.I. Budker Institute of Nuclear Physics SB RAS (BINP), Novosibirsk 630090, Russia\\
% ...
$^{14}$ Helmholtz Institute Mainz, Johann-Joachim-Becher-Weg 45, D-55099 Mainz, Germany\\
% ...
$^{20}$ (A)INFN Laboratori Nazionali di Frascati, I-00044, Frascati, Italy; (B)INFN and University of Perugia, I-06100, Perugia, Italy\\
% ...
$^{24}$ Joint Institute for Nuclear Research, 141980 Dubna, Moscow region, Russia\\
% ...
$^{39}$ Southeast University, Nanjing 211100, People's Republic of China\\
$^{40}$ State Key Laboratory of Particle Detection and Electronics, Beijing 100049, Hefei 230026, People's Republic of China\\
% ...
$^{50}$ University of Science and Technology of China, Hefei 230026, People's Republic of China\\
% ...
$^{53}$ (A)University of Turin, I-10125, Turin, Italy; (B)University of Eastern Piedmont, I-15121, Alessandria, Italy; (C)INFN, I-10125, Turin, Italy\\
% ...
$^{d}$ Also at the Novosibirsk State University, Novosibirsk, 630090, Russia\\
% ...
}
\end{center}
\vspace{0.4cm}
\end{small}
}
\noaffiliation{}
%%%%%%%%%%%%%%%%%%%%%%%%%%%%%%%%%%%%%%%%%%%%%%%%%%%%%%%%%%%%%%%%%%%%%%%%%%%%%%%%%%%%%%%%%%

\begin{abstract}

The charm semileptonic decays $D^+\to\eta e^+\nu_{e}$ and $D^+\to\eta'e^+\nu_{e}$ are studied with
a sample of $e^+e^-$ collision data corresponding to an integrated luminosity of 2.93\,fb$^{-1}$ collected at $\sqrt{s}$ = 3.773\,GeV with the BESIII detector. 
We measure the branching fractions for $D^+\to\eta e^+\nu_{e}$ to be 
$(10.74\pm0.81\pm0.51)\times10^{-4}$, and for $D^+\to\eta'e^+\nu_{e}$ to be 
$(1.91\pm0.51\pm0.13)\times10^{-4}$, where the uncertainties are statistical and systematic, 
respectively. In addition, we perform a measurement of the form factor in the decay $D^+\to\eta e^+\nu_{e}$. All the results are consistent with those obtained by the CLEO-c experiment.

\end{abstract}

%\begin{keyword}
%\end{keyword}

%\end{frontmatter}
\maketitle

%%%%%%%%%%%%%%%%%%%%%%%%%%%%%%%%%%%%%%%%%%%%%%%%%%%%%%%%%%%%%%%%
%%%%%     Introduction       Part                  %%%%%%%%%%%%%
%%%%%%%%%%%%%%%%%%%%%%%%%%%%%%%%%%%%%%%%%%%%%%%%%%%%%%%%%%%%%%%%
%\begin{multicols}{2}

\section{Introduction}\label{sec:intro}
Charm semileptonic (SL) decays involve both the $c$-quark weak decay and the strong interaction. 
In the Standard Model, the Cabibbo-Kobayashi-Maskawa (CKM) matrix~\cite{CKM} describes the mixing among the 
quark flavors in the weak decay. The strong interaction effects in the hadronic current 
are parameterized by a form factor, which is numerically calculable with Lattice Quantum Chromodynamics (LQCD). 
The differential decay rate for the charm SL decay $\dpetasl$, neglecting the positron mass, is given by
%\begin{linenomath*}
\begin{equation}
\frac{{\rm d}\Gamma(D^+\to \eta e^+\nu_{e})}{{\rm d}q^2}=\frac{G^2_F|\vcd|^2}{24\pi^3}|\vec{p}_{\eta}|^3|f_+(q^2)|^2,\label{semi-eq}
\end{equation}
%\end{linenomath*}
where $G_F$ is the Fermi constant, $\vcd$ is the relevant CKM matrix element, $\vec{p}_{\eta}$ is the momentum of the $\eta$ meson in the $D^+$ rest frame, and $f_+(q^2)$ is the form factor parametrizing the strong interaction dynamics as a function of the squared four-momentum transfer $q^2$, which is the square of the invariant mass of the $e^+$-$\nu_e$ pair.
Precise measurements of the SL decay rates provide input
to constrain the CKM matrix element $\vcd$ and to test the theoretical descriptions of the form factor. LQCD calculations of the form factor can be tested by comparing to the ones 
determined from the partial branching fraction (BF) measurements, once the CKM matrix element $\vcd$ is known.

Moreover, the mixing $\eta$-$\eta'$ or $\eta$-$\eta'$-$G$, where $G$ stands for a glueball, is of great theoretical interest, because it concerns many aspects of the underlying dynamics and hadronic structure of pseudoscalar mesons and glueballs~\cite{etamixing}. The SL decay $\dppsl$ can be used to study the $\eta$-$\eta'$ mixing in a much cleaner way than in hadronic processes due to the absence of final-state interaction~\cite{DiDonato:2011kr}. 

Based on a data sample with an integrated luminosity of  818\,pb$^{-1}$ collected
at $\sqrt{s}=3.77\gev$, the CLEO collaboration measured the BF for $\dpetasl$ and $\dpetapsl$ to be $\br{\etasl}=(11.4\pm0.9\pm0.4)\times10^{-4}$ and $\br{\etapsl}=(2.16\pm0.53\pm0.07)\times10^{-4}$~\cite{Yelton:2010js}, respectively. In this paper, we present new measurements of these BFs, using $\ddbar$ meson pairs produced near threshold at $\sqrt{s}=3.773\gev$
with an integrated luminosity of 2.93\,fb$^{-1}$~\cite{luminosity1} collected with the BESIII detector~\cite{:2009vd}.  In addition, the modulus of the form factor $f_+(q^2)$ in $\dpetasl$ is measured. 

\section{The BESIII Detector}\label{sec:BESIII}

The Beijing Spectrometer (BESIII) detects $\ee$ collisions produced by the
double-ring collider BEPCII. BESIII is a general-purpose
detector~\cite{:2009vd} with 93\,\% coverage of the full solid angle. From
the interaction point (IP) to the outside, BESIII is equipped with a
main drift chamber (MDC) consisting of 43 layers of drift cells, a
time-of-flight (TOF) counter with double-layer scintillator in the
barrel part and single-layer scintillator in the end-cap part, an
electromagnetic calorimeter (EMC) composed of 6240 CsI(Tl) crystals, a
superconducting solenoid magnet providing a magnetic field of 1.0\,T
along the beam direction, and a muon counter containing multi-layer
resistive plate chambers installed in the steel
  flux-return yoke of the magnet. The MDC spatial resolution is about
135\,$\mu$m and the momentum resolution is about 0.5\,\% for a charged
track with transverse momentum of 1\gevc. The energy resolution for
electromagnetic showers in the EMC is 2.5\,\% at 1\gev.  More details of the
spectrometer can be found in Ref.~\cite{:2009vd}.

\section{MC Simulation}\label{sec:MC}

Monte Carlo (MC) simulation serves to estimate the detection efficiencies
and to understand background components. High statistics MC samples are
generated with a {\sc geant4}-based~\cite{geant4} software
package, which includes simulations of the
geometry of the spectrometer and interactions of particles with the
detector materials. {\sc kkmc} is used to model the beam energy spread
and the initial-state radiation (ISR) in the $\ee$
annihilations~\cite{Jadach:2000ir}. The `inclusive' MC samples consist
of the production of $D\overline{D}$ pairs with consideration of
quantum coherence for all neutral $D$ modes, the
non-$D\overline{D}$ decays of $\psi(3770)$, the ISR production of low
mass $\psi$ states, and continuum processes (quantum electrodynamics (QED) and $q\bar{q}$). 
Known decays recorded by the Particle Data Group (PDG)~\cite{PDG} are
simulated with {\sc evtgen}~\cite{evtgen} and the unknown decays with
{\sc lundcharm}~\cite{lund}. The final-state radiation (FSR) of
charged tracks is taken into account with the {\sc photos}
package~\cite{photos}. The equivalent luminosity of the inclusive MC samples is about 10 times that of the data.
The signal processes of $D^{+}\to \etappsl$ are generated using the modified pole model of Ref.~\cite{Bec and Kaida}.

\section{Data Analysis}

As the $\psi(3770)$ is close to the $D\bar{D}$ threshold, the pair of $D^+D^-$ mesons is produced nearly at rest without accompanying additional hadrons. Hence, it is straightforward to use the $D$-tagging method~\cite{Baltrusaitis:1985iw} to measure the absolute BFs, based on the following equation
%\begin{linenomath*}
\begin{equation}
    \br{\etappsl} = \frac{n_{\etappsl, {\rm tag}}}{n_{\rm tag}}\cdot\frac{\varepsilon_{\rm tag}}{\varepsilon_{\etappsl, {\rm tag}}}.\label{22}
\end{equation}
%\end{linenomath*}
Here, $n_{\rm tag}$ is the total yield of the single-tag (ST) $D^-$ mesons reconstructed with hadronic decay modes, while $n_{\etappsl, {\rm tag}}$ is the number of the $\dppsl$ signal events when the ST $D^-$ meson is detected. $\varepsilon_{\rm tag}$ and $\varepsilon_{\etappsl, {\rm tag}}$ are the corresponding detection efficiencies.
Note that in the context of this paper, charge conjugated modes are always implied.

\subsection{Reconstruction of the hadronic tag modes}\label{sec:ana}

The $D^-$ decay modes used for tagging are $K^+\pi^-\pi^-, K^+\pi^-\pi^-\pi^0, \ks\pi^-, \ks\pi^-\pi^0, \ks\pi^+\pi^-\pi^-$ and $K^+K^-\pi^-$, where $\pi^0\to\gamma\gamma$, and $\ks\to\pi^+\pi^-$. The sum of the BFs of these six decay modes is about 27.7$\%$. $D^-$ tag candidates are reconstructed from all possible combinations of final state particles, according to the following selection criteria.

Momenta and impact parameters of charged tracks are measured by the
MDC. Charged tracks are required to satisfy $|\cos\theta|<0.93$, where
$\theta$ is the polar angle with respect to the beam axis, and have a
closest approach to the interaction point (IP) within $\pm10$\,cm along the beam direction
and within $\pm1$\,cm in the plane perpendicular to the beam
axis. Particle identification\,(PID) is implemented by combining the
information of specific energy loss (d$E$/d$x$) in the MDC and
the time of flight measurements from the TOF into PID likelihoods for the different particle hypotheses. For a charged $\pi$($K$)
candidate, the likelihood of the $\pi$($K$) hypothesis is required to
be larger than that of the $K$($\pi$) hypothesis.

Photons are reconstructed as energy deposition clusters in the
EMC. The energies of photon candidates must be larger than 25\,MeV for
$|\cos\theta|<0.8$ (barrel) and 50\,MeV for $0.86<|\cos\theta|<0.92$
(end cap). To suppress fake photons due to electronic noise or beam
backgrounds, the shower time must be less than 700\,ns
from the event start time~\cite{t0}.

The $\pi^0$ candidates are selected from pairs of photons of which at least one is
reconstructed in the barrel. The two photon invariant mass, $M(\gamma\gamma)$,
is required to lie in the range (0.115, 0.150)$\gevcc$. We further constrain the invariant mass of each
photon pair to the nominal $\pi^0$ mass, and update the four-momentum of the candidate according to the fit results.

The $K^0_S$ candidates are reconstructed via $K^0_S\to\pi^{+}\pi^{-}$
using a vertex-constrained fit to all pairs of oppositely charged
tracks, without PID requirements. The distance of closest approach of a charged track to the IP
is required to be less than 20\,cm along the beam direction, without requirement in the transverse plane.
The $\chi^2$ of the vertex fit is required to be
less than 100.
The invariant mass of the $\pi^{+}\pi^{-}$ pair is required to be within $(0.487,
0.511)\gevcc$, which corresponds to
three times the experimental mass resolution.

Two variables, the beam-constrained mass, $\mbc$, and the energy difference, $\delE$, are used
to identify the tagging signals, defined as follows
%\begin{linenomath*}
\begin{equation}
   \mbc\equiv\sqrt{E^2_{\unit{beam}}/c^4-|\vec{p}_{D^-}|^2/c^2},\label{25}
    %\nonumber
\end{equation}
%\end{linenomath*}
%\begin{linenomath*}
\begin{equation}
    \delE\equiv E_{D^-}-E_{\unit{beam}}.\label{26} %\nonumber
\end{equation}
%\end{linenomath*}
Here, $\vec{p}_{D^-}$ and $E_{D^-}$ are the total momentum and energy of
the $D^-$ candidate in the rest frame of the initial $e^+e^-$ system, and $E_{\unit{beam}}$ is the beam energy. Signals
peak around the nominal $D^-$ mass in $\mbc$ and around zero in
$\delE$. Boundaries of $\delE$ requirements are set at 
$\pm3\sigma$, except that those of modes containing a $\pi^0$ are set as
($-4\sigma, +3\sigma$) due to the asymmetric distributions. Here, $\sigma$ is the standard deviation from the nominal value of $\delE$. In each
event, only the combination with the least $|\delE|$
is kept per $D^-$-tagging mode.

\begin{figure*}[ht!]
\centering
\includegraphics[width=0.8\linewidth]{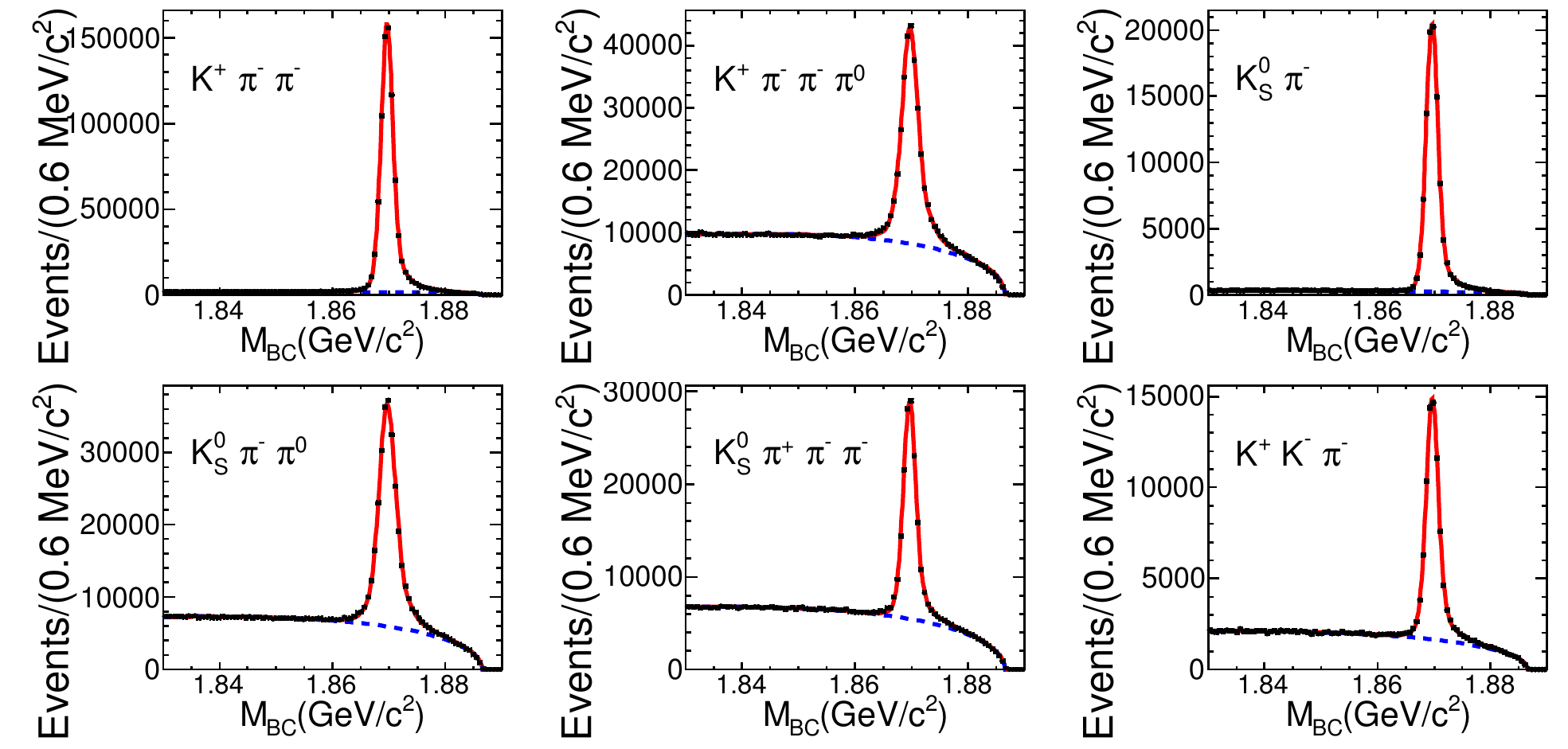}
\caption{Distributions of $\mbc$ for the six ST modes. Data are shown as points with error bars. The solid lines are
  the total fits and the dashed lines are the background
  contribution.}{\label{ST_all}}
\centering
\end{figure*}

\begin{table}[ht!]
\caption{Requirements on $\Delta E$, detection efficiencies and signal yields for the different ST modes. The errors are all statistical.}{\label{deltaE}}
\begin{tabular}{lccS[table-format=7.0(4)]}
\hline
\hline
Modes     \Tstrut   & $\delE$ ($\gev$)   & $\epsilon_{\rm tag}$ (\%) & {$n_{\rm tag}$}  \\
\hline
$K^+\pi^-\pi^-$       & $[-0.023,0.022]$  & $50.94\pm0.03$ & 801283(949) \\
$K^+\pi^-\pi^-\pi^0$  & $[-0.058,0.032]$  & $25.40\pm0.03$ & 246770(699) \\
$\ks\pi^-$            & $[-0.023,0.024]$  & $52.59\pm0.09$ & 97765(328) \\
$\ks\pi^-\pi^0$       & $[-0.064,0.037]$ & $28.07\pm0.03$ & 217816(632) \\
$\ks\pi^+\pi^-\pi^-$  & $[-0.027,0.025]$ & $32.28\pm0.05$ & 126236(425) \\
$K^+K^-\pi^-$  \Bstrut& $[-0.020,0.019]$ & $40.08\pm0.08$ & 69869(326) \\
\hline
\hline
\end{tabular}
\end{table}

After applying the $\delE$ requirements in Table~\ref{deltaE} in all the ST modes, we plot their $\mbc$ distributions in
Fig.~\ref{ST_all}. Maximum likelihood fits to these $M_{\rm BC}$ distributions
are performed, where in each mode the signals are modeled with the
MC-simulated signal shape convolved with a smearing
Gaussian function with free parameters, and the backgrounds are modeled with the ARGUS
function~\cite{Argus}. The Gaussian functions are supposed to
compensate for the resolution differences between data and MC
simulations. Based on the fit results, ST yields of data are given in Table~\ref{deltaE} in the $\mbc$ mass range [1.86, 1.88] $\gevcc$,
along with their MC-determined detection efficiencies.

\subsection{Reconstruction of SL signals}\label{sec:signal}

We look for the SL signal of $\dppsl$ in the events when the ST $D^-$ mesons are found to satisfy the requirement  $1.86 \leq \mbc \leq 1.88 \gevcc$. The positron and $\etapp $ are reconstructed from the remaining tracks and neutral clusters that have not been used in the ST $D^-$ selection. Two $\eta$ decay modes $\eta\to\gamma\gamma$ (denoted as $\eta_{\gamma\gamma}$)
and $\eta\to\pi^+\pi^-\pi^0$  (denoted as $\eta_{3\pi}$), and three $\eta'$ decay modes $\eta'\to\pi^+\pi^-\eta_{\gamma\gamma}$, $\eta'\to\pi^+\pi^-\eta_{3\pi}$
and $\eta'\to\gamma\rho^0 \to \gamma\pi^+\pi^-$, are studied. As the neutrino in the final states is undetectable at BESIII,
the SL signals are identified by studying the variable $U_{\rm{miss}}=E_{\rm{miss}}-c|\vec{p}_{\rm{miss}}|$,
where  $E_{\rm{miss}} = E_{\rm{beam}}-E_{\etapp }-E_{e^+}$  and $\vec{p}_{\rm
miss}=\vec{p}_{D^+}-\vec{p}_{\etapp}-\vec{p}_{e^+}$.
$\vec{p}_{D^+}$ is the momentum of the $D^+$ meson, $E_{\rm
\etapp}$($\vec{p}_{\etapp}$) and $E_{e^+}$($\vec{p}_{e^+}$) are
the energies (momenta) of the $\etapp$ and $e^+$,
respectively.  The momentum $\vec{p}_{D^+}$ is calculated by
$\vec{p}_{D^+}=-\hat{p}_{\rm tag}\sqrt{E_{\rm
beam}^2/c^2-m^2_{D^-}c^2}$, where $\hat{p}_{\rm tag}$ is the
momentum direction of the ST $D^-$ and
$m_{D^-}$ is the nominal $D^-$
mass~\cite{PDG}. All the momenta are calculated in the rest frame of the initial $e^+e^-$ system. 
For the signal events, the $U_{\rm miss}$ distribution is expected to peak at zero.

\begin{table*}[ht]
\caption{SL signal detection efficiencies for the different different ST  tag modes in percent. The errors are all statistical.}{\label{DTeff_all}}
\sisetup{table-format=2.2(2)}
\begin{tabular}{lSS|SSS}
\hline
\hline
Modes   \Tstrut        & \multicolumn{2}{c|}{$D^+\to\eta e^+\nu_{e}$}  &  \multicolumn{3}{c}{$D^+\to\eta'e^+\nu_{e}$} \\
Sub-decay modes             & {$\gamma\gamma$} & {$\pi^+\pi^-\pi^0$} & {$\pi^+\pi^-\eta_{\gamma\gamma}$}
& {$\pi^+\pi^-\eta_{3\pi}$} & {$\gamma\rho^0$}  \\
\hline
$K^+\pi^-\pi^-$       & 23.58(9) & 12.65(7) & 8.50(9) & 2.41(5) & 11.68(11)  \\
$K^+\pi^-\pi^-\pi^0$  & 9.77(7)  & 4.75(5)  & 3.48(6) & 0.82(3) &  4.96(7)   \\
$ \ks\pi^-$           & 25.23(9) & 13.45(8) & 9.23(9) & 2.29(5) & 12.47(11)  \\
$\ks\pi^-\pi^0$       & 9.82(7)  & 5.40(5)  & 4.60(7) & 0.83(3) &  5.83(8)   \\
$ \ks\pi^+\pi^-\pi^-$ & 13.98(8) & 6.24(5)  & 4.09(6) & 0.82(3) &  5.87(8)   \\
$K^+K^-\pi^-$ \Bstrut & 18.41(9) & 9.93(7)  & 6.28(8) & 1.52(4) &  8.18(9)   \\
%\hline
\hline
\hline
\end{tabular}
\end{table*}

Candidates for charged tracks, photons and $\pi^0$ are selected following the same selection criteria described above for the tagging $D^-$ hadronic modes.
To select the $\eta\to\gamma\gamma$ candidates, the two-photon invariant mass is required to be within (0.50, 0.58)\,$\gevcc$. A 1-C kinematic fit is
performed to constrain this mass to the nominal $\eta$ mass, and the $\chi^2$ is required to be less than 20.
If there are multiple  $\eta\to\gamma\gamma$  candidates, only the one with the least $\chi^2$ is kept. The $\eta\to\pi^+\pi^-\pi^0$ candidates are required to have an invariant mass within (0.52, 0.58)\,$\gevcc$. If multiple candidates exist per event, we only keep the candidate closest to the nominal $\eta$ mass. 
In the reconstruction of $\dpetapsl$ signals, $\eta'\to\pi^+\pi^-\eta$ candidates are formed by combining an $\eta$ candidate with two charged pions. Their invariant mass must lie in (0.935, 0.980)\,$\gevcc$ for $\etap\to\pi^+\pi^-\eta_{2\gamma}$ and  in (0.930, 0.980)\,$\gevcc$ for $\etap\to\pi^+\pi^-\eta_{3\pi}$; if multiple candidates are found, only the one closest to the nominal $\etap$ mass is chosen. For $\etap \to\gamma\rho^0$ candidate, we require a mass window (0.55, 0.90)\,$\gevcc$ for $\rho^0\to\pi^+\pi^-$ candidates, and the radiative photon is not to form a $\pi^0$ candidate with any other photon in the event. The energy of the radiative photon is required to be larger than 0.1\,$\gev$ in order to suppress $D^+\to\rho^0 e^+\nu_{e}$ backgrounds. The helicity angle of the daughter pion in the rest frame of $\rho^0$, $\theta_{\pi\rho}$, is required to satisfy $|\cos\theta_{\pi\rho}|<0.85$.
To suppress backgrounds from FSR, the angle between the direction of the radiative photon and the positron momentum is required
to be greater than 0.20 radians. Furthermore, the angles between the radiative photon and all charged tracks in the final state of the $D^-$ tag candidates are
required to be larger than 0.52 radians, to suppress fake photons due to split-offs from hadronic showers in the EMC.

The positron is tracked in the MDC and distinguished from other charged 
particles by combining the d$E$/d$x$, TOF and EMC information.
 The determined PID likelihood $\mathcal{L}$ is required to satisfy $\mathcal{L}(e)>0$ and  $\mathcal{L}(e)/(\mathcal{L}(e)+\mathcal{L}(\pi) + \mathcal{L}(K))>0.8$. Furthermore, the energy measured in the EMC divided by the track momentum is required
to be larger than 0.8 for $\dpetasl$ and larger than 0.6 for $\dpetapsl$.  In addition,  positron candidates with momentum less than $0.2\gevc$ are discarded in $\dpetapsl$ decays to reduce mis-PID rate. Events that have extra unused EMC showers with energies larger than 250~MeV,
 are discarded.

\begin{figure*}[ht!]
\includegraphics[width=0.8\linewidth]{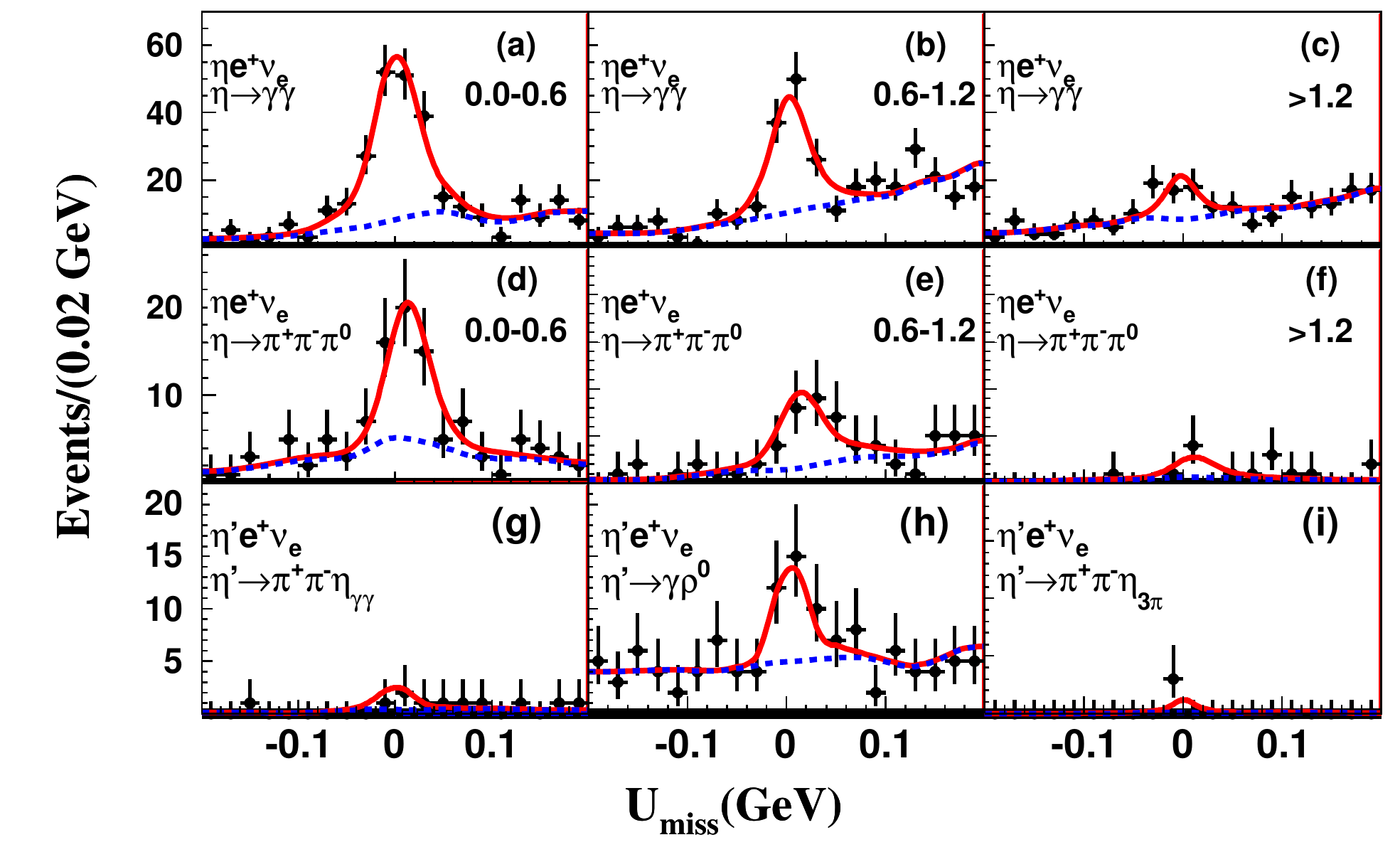}
\caption{Distributions of $U_{\rm {miss}}$ for the different signal modes. Data are shown as points with error bars. The solid lines are
  the total fits and the dashed lines are the background
  contributions. Data for $\dpetasl$ are plotted in 3 bins of 0.0$\leq q^2<$0.6 GeV$^2/c^4$ (a, d), 0.06$\leq q^2 \leq $1.2 GeV$^2/c^4$ (b, e) and $q^2>1.2$ GeV$^2/c^4$ (c, f).  }{\label{DT_all}}
\end{figure*}

The resultant $U_{\rm miss}$ distributions are plotted in Fig.~\ref{DT_all}. We perform simultaneous unbinned maximum likelihood fits to the different decay modes for $\etasl$ and $\etapsl$, respectively. The signal shapes are obtained from MC simulations convolved with Gaussian functions whose widths are determined from the fit to account for the resolution difference in data and MC. The widths are around $15\%$ of the total resolution. The background shapes of different $\etapp$ decay modes are modeled with the distributions from backgrounds obtained from the inclusive MC sample. In total, we observe $373\pm26$ signal events for $\dpetasl$ and $31.6\pm8.4$ for $\dpetapsl$. The BF for $\dppsl$ is determined by using Eq.~\eqref{22} according to the MC-determined efficiencies in Table~\ref{DTeff_all}, which gives $\br{\eta e^+\nu_{e}}=(10.74\pm0.81)\times10^{-4}$,
and $\br{\eta'e^+\nu_{e}}=(1.91\pm0.51)\times10^{-4}$.
 
The statistics of $D^+\rightarrow\eta e^{+}\nu_{e}$ allows to determine $|f_+(q^2)|$, as defined in Eq.~\eqref{semi-eq}. Hence, a fit is implemented to the partial BFs in the three $q^{2}$ bins used in Fig.~\ref{DT_all}. By introducing the life time $\tau_{D^+}$=(1040$\pm$7)$\times10^{-15}s$ from PDG~\cite{PDG}, we construct  
$\chi^{2}=\Delta\gamma^{T}V^{-1}\Delta\gamma$, where $\Delta\gamma=\Delta\Gamma_{m}-\Delta\Gamma_{p}$ 
is the vector of differences between the measured partial decay widths $\Delta\Gamma_{m}$ and the expected partial 
widths $\Delta\Gamma_{p}$ integrated over the different $q^2$ bins, and $V$ is the total covariance matrix consisting of the statistical covariance matrix $V_{\rm stat}$ 
and the systematic covariance $V_{\rm syst}$. The statistical correlations among the different $q^2$ bins are negligible. We list the elements of the total covariance matrix $V$ in Table~\ref{cmatrix}.

\begin{table}[htbp]
\caption{Correlation matrix including statistical and systematic contributions in the fit.}{\label{cmatrix}}
%\begin{center}
\begin{ruledtabular}
\begin{tabular}{lccc}
%\hline
%\hline
$q^2(\rm GeV^2/c^4)$ & $0.0-0.6$  & $0.6-1.2$ & $>1.2$   \\
\hline
  $0.0-0.6$      & $1$          & $0.075$    & $0.032$\\
$0.6-1.2$   	 & $0.075$		& $1$	     & $0.026$\\
$>1.2$ 			 & $0.032$		& $0.026$	 & $1$\\
%\hline
%\hline
\end{tabular}
\end{ruledtabular}
%\end{center}
\end{table}

Three parameterizations of the form factor $f_+(q^2)$ are adopted in the fits.  The first form is the simple pole model of Ref.~\cite{Bec and Kaida}, which is given as
%\begin{linenomath*}
\begin{align}
f_+(q^2) &= \frac{f_+(0)}{1-\frac{q^2}{m_{\rm pole}^2}}.
\end{align}
%\end{linenomath*}
Here, $m_{\rm pole}$ is predicted to be close to the mass of $D^{*+}$~\cite{PDG}, which is 2.01 $\gevcc$ and is a free parameter in the fit. The second choice is the modified pole model~\cite{Bec and Kaida}, written as
%\begin{linenomath*}
\begin{align}
f_+(q^2)&= \frac{f_+(0)}{(1-\frac{q^2}{m_{\rm pole}^2})(1-\alpha\frac{q^2}{m_{\rm pole}^2})}, 
\end{align}
%\end{linenomath*}
where $m_{\rm pole}$ is fixed at the mass of $D^{*+}$ and $\alpha$ is a free parameter to be determined.
The third is a general series parametrization with $z$-expansion, which is formulated as
%\begin{linenomath*}
\begin{align}
f_+(q^2) &= \frac{1}{P(q^2)\phi(q^2, t_0)}\sum_{k=0}^\infty a_k(t_0){[z(q^2, t_0)]}^k.
\end{align}
%\end{linenomath*}
Here, $t_0=t_+(1-\sqrt{1-t_-/t_+})$ with $t_\pm=(m_{D^+}\pm m_\eta)^2$ and $a_k(t_0)$ are real coefficients.
The functions $P(q^2)$, $\phi(q^2, t_0)$ and $z(q^2, t_0)$ are formulated following the definitions in Ref.~\cite{Z-expan}.
In the fit, the series is truncated at $k=1$.

\begin{figure}[tp!]
\centering
\includegraphics[width=0.8\linewidth]{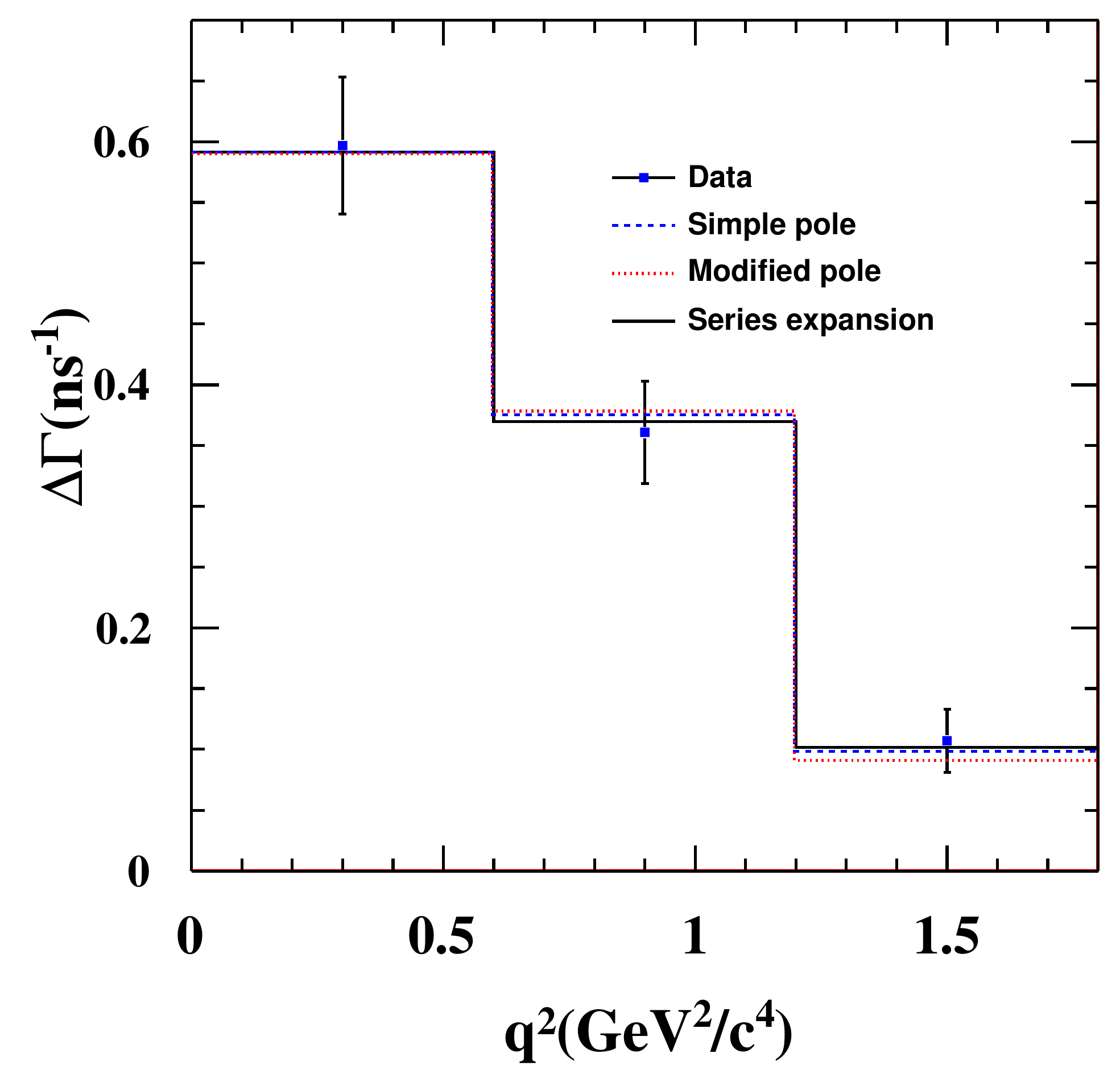}
\caption{Fit to the partial widths of $\dpetasl$. The dots with error bars are data and the lines are the fits with different form-factor models.}{\label{fffit}}
\centering
\end{figure}

Three separate fits to data are implemented, based on the three form-factor models. Their fit curves are plotted in Fig.~\ref{fffit}. We determine the values of $f_{+}(0)|V_{cd}|$ in all three scenarios, as listed in Table~\ref{fffit_results}. We observe that the results of $f_{+}(0)|V_{cd}|$ in the three fits are consistent and the fit qualities are good.

\begin{table*}[htbp]
\caption{The fit results of the form-factor parameters. For simple pole and modified pole parameterizations, shape parameters denote $m_{\rm pole}$ and $\alpha$, respectively. For the series parametrization, we provide results of $f_{+}(0)|V_{cd}|$, $r_1=a_1/a_0$ (shape parameter). The correlation coefficients $\rho$ between fitting parameters and the reduced $\chi^2$ are given.}{\label{fffit_results}}
%\begin{center}
\begin{ruledtabular}
\begin{tabular}{lccc}
%\hline
%\hline
Fit parameters & Simple pole  & Modified pole & Series expansion   \\
\hline
$f_+(0)|V_{cd}|$ ($\times10^{-2}$)          & $8.15\pm0.45\pm0.18$    & $8.24\pm0.51\pm0.22$      & $\phantom{-}7.86\pm0.64\pm0.21$       \\
Shape parameter   	   & $1.73\pm0.17\pm0.03$		&$0.50\pm0.54\pm0.08$			&$-7.33\pm1.69\pm0.40$\\
$\rho$ 					&0.80				&$-0.85$			&0.90\\
$\chi^2/\rm{ndf}$				& $0.1/(3-2)$				&$0.3/(3-2)$	& $0.5/(3-2)$\\
%\hline
%\hline
\end{tabular}
%\end{center}
\end{ruledtabular}
\end{table*}

%\newpage

\section{Systematic Uncertainties}\label{sysu}

With the double-tag technique, the systematic uncertainties in detecting the ST $D^-$ mesons in the BF measurements mostly cancel as shown in Eq.~\eqref{22}. For the SL signal side, the following sources of systematic uncertainties are studied, as summarized in Table~\ref{sys_br}. All of these contributions are added in quadrature to obtain the total systematic
uncertainties on the BFs. 

The uncertainties of tracking and PID
efficiencies for $\pi^{\pm}$ are studied with control samples of $\ddbar$ Cabibbo favored ST decays~\cite{dkorpienu}. The uncertainties in $e^{\pm}$
tracking and PID efficiencies are estimated with radiative Bhabha events, taking account of the different tracking and PID efficiencies in different $\cos\theta$ and momentum distributions of $e^{\pm}$.

The uncertainty due to the $\pi^0$ and $\eta$ reconstruction efficiency is estimated with a control sample using $D^0\to K^-\pi^+\pi^0$ selected without requiring the $\pi^0$ meson. The uncertainties associated with the $\eta$ and $\eta'$ invariant mass requirements are estimated by changing the
requirement boundaries and taking the maximum variations of the resultant BFs as systematic uncertainties. The uncertainty due to the extra shower veto is studied with doubly tagged hadronic events, and is found to be negligible.

The uncertainties of the radiative $\gamma$ selection in $\eta'\to\gamma\rho^0$ are studied using a control sample from $D^0\bar{D}^0$ decays where the $D^0$ meson decays to $\ks\eta', \eta'\to\gamma\rho^0$ and the $\bar{D}^0$ decays to Cabibbo favored ST modes. We impose the same selection criteria on the radiative photon to the control sample, and the difference of signal survival rates between data and MC simulations is found to be $3.1\%$.  The uncertainty due to the $\rho$ invariant mass requirement is also estimated with this control sample. The difference of signal survival rates between data and MC simulations is found to be $0.6\%$.

In the fit to the $U_{\rm miss}$ distribution, the uncertainty due to the parametrization of the signal shape is
estimated by introducing a Gaussian function to smear the MC-simulated signal shape and varying the parameters of the smearing Gaussian. The uncertainty due to the background modeling is estimated by changing the background model to a 3rd degree Chebychev polynomial. The uncertainty due to the fit range is estimated by repeating the fits in several different ranges. The uncertainties of the input BFs and the limited MC statistics are also taken into account.

We also study the $\delE$ and $\mbc$ requirements by varying the ranges and compare the efficiency-corrected tag yields. The resultant maximum differences are taken as systematic uncertainties. 
The SL signal model for $\dpetasl$ is simulated according to the form factor measured in this work and the variations within one standard deviation are studied. For $\dpetapsl$, since there is no available form-factor data, we take the form factor of $\dpetasl$ and evaluate the systematic uncertainty as we do for $\dpetasl$.

\begin{table}[htbp]
\caption{Relative systematic uncertainties in the BF measurements (in $\%$). The lower half of the table presents the common uncertainties among the different channels.}{\label{sys_br}}
\begin{ruledtabular}
\begin{tabular}{lcc|ccc}
Source                & \multicolumn{2}{c|}{$D^+\to\eta e^+\nu_{e}$}  &  \multicolumn{3}{c}{$D^+\to\eta'e^+\nu_{e}$} \\
Sub-decay modes             & $\gamma\gamma$ & $\pi^+\pi^-\pi^0$ & $\pi^+\pi^-\eta_{\gamma\gamma}$ 
& $\pi^+\pi^-\eta_{3\pi}$ & $\gamma\rho$  \\
\hline
$\pi^\pm$ tracking and PID                 &  & 2.8 & 4.1 & 8.2 & 1.6  \\
$\pi^0/\eta$ reconstruction    & 2.0 & 2.0 & 2.2  & 2.2 &  \\
Input BF   & 0.3 & 0.3 & 1.7  & 2.0 & 1.7  \\
$\rho$ mass window & & & & &0.6\\
Radiative $\gamma$    &  &  &   &  & 3.1  \\
$\eta'$ mass window            & \multicolumn{2}{c|}{}  & 1.8  & 1.6 & 1.9  \\ \hline
$e^+$ tracking and PID           & \multicolumn{2}{c|}{1.1}   & \multicolumn{3}{c}{3.7}    \\
$\eta$ mass window             & \multicolumn{2}{c|}{2.4}   & \multicolumn{3}{c}{2.4}    \\
$U_{\rm miss}$ fit                        & \multicolumn{2}{c|}{2.1}  &  \multicolumn{3}{c}{1.0}   \\
$\delE/\mbc$ window                       & \multicolumn{2}{c|}{0.9}           & \multicolumn{3}{c}{0.9}  \\
MC statistics					& \multicolumn{2}{c|}{0.2}  & \multicolumn{3}{c}{0.5}\\
SL signal model		& \multicolumn{2}{c|}{0.9} & \multicolumn{3}{c}{0.9}\\
\hline
Total                            & \multicolumn{2}{c|}{4.7}  & \multicolumn{3}{c}{6.9}   \\
\end{tabular}
\end{ruledtabular}
\end{table}

Systematic uncertainties of the partial decay widths of $\dpetasl$ to calculate $V_{\rm syst.}$ are studied following the same procedure mentioned above. For most of the common systematics, we quote the values from the total BF measurements in Table~\ref{sys_br}.
For charged pion tracking 
and PID, we evaluate the uncertainty averaged over the two $\eta$ decay modes according to their relative yields. 
For $e^+$ tracking and PID, we reweight the systematic uncertainties in each $q^2$ bin. All these items are summarized in Table~\ref{sys_err_patialdecayrate}. For the systematics of $\eta$ mass window and fitting procedure, we refit the $U_{\rm miss}$ distribution after varying the $\eta$ mass window and changing fitting region and compare the refitting results of the form factors. The maximum deviations from the nominal results are calculated to be $1.3\%$ and $0.4\%$ for the $f_+(0)|V_{cd}|$ and shape parameter and are considered as systematic uncertainties. The sum of the systematic uncertainties is given in Table~\ref{fffit_results}.

\begin{table}[htbp]
\caption{Relative systematic uncertainties (in $\%$) of the measured partial decay widths of $\dpetasl$ used to obtain $V_{\rm syst.}$.  }{\label{sys_err_patialdecayrate}}
\begin{ruledtabular}
\begin{tabular}{lcccccc}
\diagbox{Source}{$q^2$ ($\gevccs$)}  & $0.0-0.6$  & $0.6-1.2$  & $>$1.2	\\ \hline
$e^+$ tracking and PID                  &1.4 &0.9 & 0.1    \\
\hline
$\pi^\pm$ tracking and PID                 & \multicolumn{3}{c}{1.7}             \\
$\pi^0/\eta$ reconstruction         & \multicolumn{3}{c}{2.0}                      \\
$\delE/\mbc$ window                        & \multicolumn{3}{c}{0.9}           \\
MC statistics					& \multicolumn{3}{c}{0.2}			\\
SL signal model		& \multicolumn{3}{c}{0.9} \\
Input BF  & \multicolumn{3}{c}{0.3} \\
$D^+$ lifetime & \multicolumn{3}{c}{0.7} \\
\hline
Total		 &3.3 &3.0  &2.9\\	
\end{tabular}
\end{ruledtabular}
\end{table}

\section{Summary}
We exploit a double-tag technique to analyze a sample of 2.93\,fb$^{-1}$ $\ee\to D^{+}D^{-}$ at $\sqrt{s}=3.773$ GeV. The BF for the SL decay
$\dpetasl$ is measured to be $\br{\eta e^+\nu_{e}}=(10.74\pm0.81\pm0.51)\times10^{-4}$,
and for $D^+\to\eta'e^+\nu_{e}$ to be $\br{\eta'e^+\nu_{e}}=(1.91\pm0.51\pm0.13)\times10^{-4}$,
where the first and second uncertainties are statistical and systematic, respectively.
In addition, we measure the decay form factor for $D^+\to\eta e^+\nu_{e}$ based on three form-factor models, whose results are given in Table~\ref{fffit_results}. This helps to calibrate the form-factor calculation in LQCD.
All these results are consistent with the previous measurements from CLEO-c~\cite{Yelton:2010js}. 
Our precision is only slightly better than CLEO-c's, because our limitations on PID and low-momentum tracking efficiency hinder to adopt CLEO-c's generic $D$-tagging method~\cite{Yelton:2010js}.
We average the results of $\br{\eta e^+\nu_{e}}$ and $\br{\eta'e^+\nu_{e}}$ in the two experiments to be $(11.04\pm0.60\pm0.33)\times10^{-4}$ and $(2.04\pm0.37\pm0.08)\times10^{-4}$, respectively. Using the input value recommended by Ref.~\cite{etamixing}, the $\eta-\eta'$ mixing angle $\phi_{P}$ is determined to be $(40\pm3\pm3)\degree$, where the first uncertainty is experimental and the second theoretical, in agreement with the results obtained by Ref.~\cite{etamixing}. However, the current precision for $\dppsl$ is not enough to provide meaningful constraints on the $\eta$-$\eta'$ mixing parameters.

%\newpage
%%%%%%%%%%%%%%%%%%%%%%%%%%%%%%%%%%%%%%%%%%%%%%%%%%%%%%%%%%%%%%%%
%%%%%    acknowledgments       Part                %%%%%%%%%%%%%
%%%%%%%%%%%%%%%%%%%%%%%%%%%%%%%%%%%%%%%%%%%%%%%%%%%%%%%%%%%%%%%%
\begin{acknowledgments}
The BESIII collaboration thanks the staff of BEPCII, the IHEP computing center and the supercomputing center of USTC for their strong support. This work is supported in part by National Key Basic Research Program of China under Contract No. 2015CB856700; National Natural Science Foundation of China (NSFC) under Contracts Nos. 11605198, 11335008, 11375170, 11425524, 11475164, 11475169, 11605196, 11625523, 11635010, 11735014; the Chinese Academy of Sciences (CAS) Large-Scale Scientific Facility Program; the CAS Center for Excellence in Particle Physics (CCEPP); Joint Large-Scale Scientific Facility Funds of the NSFC and CAS under Contracts Nos. U1532102, U1532257, U1532258, U1732263; CAS Key Research Program of Frontier Sciences under Contracts Nos. QYZDJ-SSW-SLH003, QYZDJ-SSW-SLH040; 100 Talents Program of CAS; INPAC and Shanghai Key Laboratory for Particle Physics and Cosmology; German Research Foundation DFG under Contracts Nos. Collaborative Research Center CRC 1044, FOR 2359; Istituto Nazionale di Fisica Nucleare, Italy; Koninklijke Nederlandse Akademie van Wetenschappen (KNAW) under Contract No. 530-4CDP03; Ministry of Development of Turkey under Contract No. DPT2006K-120470; National Science and Technology fund; The Swedish Research Council; U. S. Department of Energy under Contracts Nos. DE-FG02-05ER41374, DE-SC-0010118, DE-SC-0010504, DE-SC-0012069; University of Groningen (RuG) and the Helmholtzzentrum fuer Schwerionenforschung GmbH (GSI), Darmstadt.
\end{acknowledgments}

%\section{References}

\end{document}